\documentclass[12pt]{article}

\input epsf
\usepackage{amssymb,wrapfig}
\usepackage{xypic,multirow}
\usepackage[matrix,arrow,curve]{xy}
\usepackage{cite}

\makeatletter
\@addtoreset{equation}{section}
\makeatother

\def\nn {{\cal N}}
\def\rr {{\Bbb R}}
\def\cc {{\Bbb C}}

\def\del {\partial}
\def\cy {Calabi--Yau}
\def\ka {K\"ahler}
\def\del {\partial}
\def\tts {{$T\oplus T^*$}}
\def\stt {{$\mathrm{SU(3)}\times\mathrm{SU(3)}$}}

\makeatletter
\@addtoreset{equation}{section}
\makeatother

\begin{document}
	
	\begin{titlepage}
	\begin{center}

	                                \hfill   SU-ITP-07/06\\
	\vskip .3in \noindent


	{\Large \bf{Reformulating Supersymmetry\\\vspace{.1cm}  with a Generalized Dolbeault 
	Operator}}

	\bigskip

	 Alessandro Tomasiello\\

	\bigskip

	ITP, Stanford University, Stanford CA 94305-4060, USA


	\vskip .5in
	{\bf ABSTRACT }
	\vskip .1in
	\end{center}

	\noindent The conditions for $\nn=1$ supersymmetry in type II supergravity have been 	 
	previously reformulated in terms of generalized complex geometry. We improve that 
	reformulation so as to completely eliminate the remaining explicit dependence on
	the metric. Doing so involves a natural generalization of the Dolbeault operator.
	As an application, we present some general arguments about supersymmetric 
	moduli. In particular, a subset of them are then classified by a certain cohomology.
	We also argue that the Dolbeault reformulation should make it easier to find existence  
	theorems for the $\nn=1$ equations.
	\vfill
	\eject


	\end{titlepage}

\section{Introduction}\label{sec:introduction} 

Even if we know that supersymmetry is not realized in our vacuum, it might well be spontaneously
broken at energy scales well below the Planck scale. It seems a good idea, then, to understand 
the structure of the supersymmetric vacua of supergravity, as a first step before 
implementing  spontaneous supersymmetry breaking. 

It might seem that there are already, if anything, too many supersymmetric vacua in string 
theory. The vast majority of those considered so far are, however, based on \cy\ manifolds;
among flux vacua, most are special cases of the same type of construction
\cite{dasgupta-rajesh-sethi,grana-polchinski,becker2}. One of the possible dangers of this lack 	
of ``genetic diversity" is that one might draw unjustified conclusions about what type of 
theories  string theory can and cannot realize at low energies. From other side, learning more 
about the general properties of geometric flux compactifications will likely teach us more about
the dynamical properties of string theory, or at least provide a more diversified testing ground.
Last but not least, going beyond \cy\ manifolds is necessary to apply AdS/CFT to many theories 
\cite{lunin-maldacena,minasian-petrini-zaffaroni,wijnholt}. 

 $\nn=1$ supersymmetry has been reformulated as a geometrical condition in \cite{gmpt2}. It is
equivalent, namely, to equations (\ref{eq:gmpt2},\ref{eq:gmpt2i}) below. The first in (\ref{eq:gmpt2}), in particular, had been studied already in \cite{hitchin-gcy,gualtieri}: it defines
what is called a generalized complex manifold (with holomorphically trivial canonical 
bundle\footnote{This is actually called generalized \cy\ in \cite{hitchin-gcy}; but this name has
become confusing after a similar--sounding definition in \cite{gualtieri}, and especially after
the somehow stronger definition of generalized \ka\ manifolds there.}).
 
The differential forms $\Phi_{1,2}$, given the algebraic
constraints that they have to satisfy (reviewed in Appendix \ref{sec:review}), determine a 
metric $g$. This metric does not appear explicitly, because (\ref{eq:gmpt2},\ref{eq:gmpt2i})
are written purely in terms of exterior calculus, which is indeed one of their attractive features. There is one place, though, where
$g$ appears explicitly, and it is the Hodge $*$ in (\ref{eq:gmpt2i}). It might naively seem that one can eliminate this $*$  by redefining $F$, but this would make it reappear in the Bianchi identity (\ref{eq:bianchi}). 
 
In section \ref{sec:susy}, we will show that we can eliminate the Hodge $*$ by reformulating the
first equation in (\ref{eq:gmpt2i}) as
\begin{equation}\label{eq:main}
	F=-8i\,(\bar\del_H^{{\cal J }_1}-\del_H^{{\cal J }_1})(e^{-3A}{\rm Im}\Phi_2)\ 
\end{equation}
where  $\del_H^{{\cal J }_1}$ is a generalization of the Dolbeault operator, 
associated in an algebraic way to 
$\Phi_1$ and to the NS three--form $H$. 
For example, when $\Phi_1$ is a holomorphic three--form $\Omega$ and $H=0$, $\del_H^{{\cal J}_1}$
is the usual Dolbeault operator $\del$. 

The point of (\ref{eq:main}) is that, while $*$ depends on the full metric, and hence
on both $\Phi_{1,2}$, the generalized Dolbeault depends only on $\Phi_1$ and not on $\Phi_2$. 
Hence, while (\ref{eq:gmpt2i}) was highly nonlinear, (\ref{eq:main}) is linear in both $\Phi_{1,2}$. 

One consequence of this is that the moduli problem for (\ref{eq:main}) becomes easier. 
It is well--known that the moduli space of Calabi--Yau compactifications without flux has a dimension determined by cohomology groups of the \cy. Once fluxes are introduced as in \cite{dasgupta-rajesh-sethi,grana-polchinski,becker2}, one can lift all of them  but the overall
volume (that in turn one can then fix by quantum effects \cite{kklt}). The equations for 
${\cal N}=1$ supersymmetry that we have just reviewed seem to have no particular preference for \cy\ manifolds. The explicit study in \cite{gmpt3} seems to indicate that more general types
of vacua are at least as abundant.

It would be interesting, then, to see if in the general case the moduli are given by some kind of generalized cohomology. Even if the non--\cy\ examples are still few, knowing the answer to such
a question would tell us what manifolds are more promising (which ones, for examples, have
already classically few or no moduli) before we even try to show the existence of a vacuum on 
them.

We attack the moduli problem (using (\ref{eq:main})) in section \ref{sec:moduli}. We do not solve it in full generality, but we give a kind of cohomological description (in (\ref{eq:nonintcoh}) below) for a subset of moduli, those coming from deformations of $\Phi_2$. We propose some speculation for
the $\Phi_1$ moduli in section \ref{sub:are}. Notice that, were it for the first equation in 
(\ref{eq:gmpt2}) alone, there would exist already a cohomological description for moduli, that
we review from several different angles in section \ref{sub:gcmoduli}. 

(\ref{eq:main}) is also likely to be a step forward towards finding an existence theorem for
supersymmetric vacua; we make this case in section \ref{sec:method}.


\section{Supersymmetry and the $d^{\cal J }$ operator}\label{sec:susy} 

In this section, we will reformulate the conditions for ${\cal N }=1$ supersymmetry
obtained in \cite{gmpt2}. If the spacetime is taken to be Minkowski, the supersymmetry equations read\footnote{Notice the different normalization of $||\Phi_{1,2}||$ with respect to \cite{gmpt2}; also,
the conventions for the RR flux in IIA have been harmonized with the ones in
IIB, as in \cite{martucci-smyth,gmpt2}.}
\begin{eqnarray}
	\label{eq:gmpt2}
	&d_H \Phi_1=0 \ , \qquad\qquad& d_H (e^{-A}{\rm Re}\Phi_2)=0 \ ,   \\
	\label{eq:gmpt2i}
	&d_H(e^A{\rm Im} \Phi_2)=\frac{e^{4A}}8 *\lambda(F) \ ,\qquad\qquad&  ||\Phi_1||=||\Phi_2||=
		\frac{e^{3A-\phi}}{\sqrt{8}}
\end{eqnarray}
Here, $\lambda$ is just some flipping of signs defined in (\ref{eq:lambda}); the norm $||\cdot||$ is defined\footnote{The equal norm condition $||\Phi_1||=||\Phi_2||$ is actually, in general, an assumption; it is automatically satisfied in the compact case (because of the
presence of an orientifold), and whenever the solution admits supersymmetric probe 
branes \cite{martucci}.} in (\ref{defMukai}). 
$d_H=d-H\wedge$ is a differential, in that it squares to zero (due to $dH=0$). $F$ is the internal flux, which via self--duality determines the whole ten--dimensional flux:
$F^{(10)}= F + {\rm vol}_4 \wedge \lambda(*F)$. $A$ is the warping, defined by $g_{10}= 
e^{2A} g_4 + g_6 $.  
$\Phi_{1,2}$ are two differential forms (of mixed degree) that define an \stt\
 structure on \tts: namely, they are compatible pure spinors\footnote{Some aspects of
generalized complex geometry are reviewed in appendix \ref{sec:review}.}.  They have different parity in IIA and IIB: 
\begin{equation}
	\label{eq:formparity}
	{\rm IIA}: \ \begin{array}{cc}\vspace{.2cm}
			\Phi_1=\Phi_+\\ \Phi_2=\Phi_-
	\end{array}
	\qquad \qquad 
	{\rm IIB}: \ \begin{array}{cc}\vspace{.2cm}
			\Phi_1=\Phi_-\\ \Phi_2=\Phi_+
	\end{array}
\end{equation}
Notice that in (\ref{eq:gmpt2}) the dilaton is decoupled from the other fields. One can
solve the system with $||\Phi_1||=||\Phi_2||$, and then, at the very end, extract the dilaton 
from in (\ref{eq:gmpt2i}). 

By their algebraic characterization, $\Phi_\pm$ define
a metric $g$ and a $B$ field. Since the equations are written in terms of
wedge products and exterior differential of forms, the metric does not appear
explicitly but through $\Phi_\pm$ -- and also, crucially, through the Hodge $*$ operator. 

For many applications, the appearance of this $*$ is annoying because $F$ is
constrained by the Bianchi identity: 
\begin{equation}
	\label{eq:bianchi}
	d_H F= \delta
\end{equation}
where $\delta$ is a magnetic source determined by the orientifolds and branes on the manifold. Putting together (\ref{eq:bianchi}) and (\ref{eq:gmpt2i}) one gets a second--order equation containing a Laplacian. Even worse, the metric
in the definition of this Laplacian is related to the form the Laplacian is acting on. 
(We will see this more explicitly in section \ref{sec:method}.)
Notice that an electric RR source is instead forbidden: $d_H (e^{4A}*\lambda(F))=0$ follows from (\ref{eq:gmpt2i}). An NSNS 
source is also forbidden, since
\begin{equation}\label{eq:d*H}
d(e^{4A-2 \phi}*H)= - e^{4A}\sum_k *F_{k+2}\wedge F_k	\ 
\end{equation}
is implied by the supersymmetry equations\footnote{This was
shown after the first version of this paper.} \cite{koerber-tsimpis,koerber-martucci-AdS}. 
 
After a review of some algebraic facts in subsection \ref{sub:the_action}, 
we will show in subsection \ref{sub:new} how to get rid of the $*$ in (\ref{eq:gmpt2i}). 

\subsection{The action of ${\cal J }$}\label{sub:the_action} 

Recall that one can associate to a pure spinor $\Phi$ a generalized almost complex structure ${\cal J }$. This is a matrix in ${\rm End}(T\oplus T^*)$ such that
\begin{equation}
	\label{eq:calJ}
{\cal J }^2=-1\ , \qquad {\cal I  }{\cal J }=-{\cal J }^t{\cal I }\ ,	
\end{equation}
where ${\cal I }={{0 \ 1}\choose{1 \ 0}}$ is the natural metric on \tts. 

In dimension 6, the fibre of \tts\ has dimension 12: a basis is given by 
the vectors $\del_m$ and the
one--forms $dx^m$. They act naturally on the bundle of differential forms 
$\Lambda^* T^*$, respectively by contraction, $\iota_{\del_m}$, and by wedge, $dx^m\wedge$. 
Let us denote their action collectively by $\Gamma_A$, $A=1\ldots 12$.

Given a matrix $q^A{}_B$ in ${\rm End}(T\oplus T^*)$ in the Lie algebra 
${\rm o}(6,6)$ (namely, one that preserves the metric ${\cal I }$ the way
${\cal J }$ does in (\ref{eq:calJ})), one can determine a natural action $q\cdot$ on the bundle of differential forms $\Lambda^* T^*$ via a Bogolubov--type computation: 
\begin{equation}
	\label{eq:bog}
	e^{q\cdot} \Gamma_A e^{-q\cdot}= 
	[\exp(q)]^A{}_B \Gamma^B\ , \qquad
	q\cdot=\frac12 q_{AB}\Gamma^{AB}\ 
\end{equation}
where we have used the metric ${\cal I }$ to lower one index: $q_{AB}= I_{AC}q^C{}_B$. Notice that $q_{AB}$ is antisymmetric precisely because
$q^A{}_B$ is in ${\rm o}(6,6)$. 

We can now apply this to $q={\cal J }$. We obtain the action 
${\cal J }\cdot\equiv \frac12 {\cal J }_{AB}\Gamma^{AB}$ \cite{cavalcanti} on the bundle of forms. Explicitly: 
\begin{equation}
	\label{eq:Jaction}
	{\cal J}\cdot= \frac12 \Big(J_{mn}dx^m\wedge dx^n \wedge +2 I^m{}_n[dx^n\wedge,\iota_{\del_n}]+ P^{mn}\iota_{\del_m}\iota_{\del_n}\Big)
	\ , \qquad 
	\left(\begin{array}{cc}
		I& P\\J&-I^t 
	\end{array}\right)\equiv {\cal J}\ .
\end{equation}

Let us see what this action is like in the two most popular examples
of ${\cal J}$. The first is one induced by a complex structure $I$ through 
${\cal J}={{I \ \ \ 0}\choose {0\ -I^t}}$. In this case, the action is just
$I^m{}_n (dx^n\wedge) \iota_{\del_m}$, which gives $i(p-q)$ on a $(p,q)$ form. 
The second case is ${\cal J}={\hspace{-.1cm}{\ \ 0\ -J^{-1}}\choose{ J \ \ \ 0}}$, with $J$ a 
symplectic two--form. In this case ${\cal J}\cdot=J\wedge - J^{-1}\llcorner$. 
Note that $J^{-1}$ is really just defined as the inverse of $J$, and without any
metric. In fact, this is a general feature of (\ref{eq:Jaction}): nothing in ${\cal J}\cdot$ requires
the metric for its definition, since $dx^m\wedge$ are only needed with their index up, 
and $\iota_{\del_n}$ with their index down. 

One nice property of this action is that its eigenforms are known.  Recall that 
the correspondence between $\Phi$ and ${\cal J }$ gives that the annihilator $L_\Phi$
of $\Phi$ is the $i$--eigenvalue of ${\cal J }$. Then the conjugate $\bar L_\Phi$ consists of creators, and by acting with its elements one obtains a basis for the bundle of differential forms. If one defines $U^3_{\cal J}$ to be generated by $\Phi$, let $U^{3-k}_{\cal J}$ 
to be obtained by the action of $k$ elements of $\bar L_\Phi$ on $\Phi$. 

It is then easy to see \cite{cavalcanti} that ${\cal J }\cdot U^{k}_{\cal J}= ki \, U^k_{\cal J} $. One first
computes $[{\cal J\cdot }, l_A \Gamma^A ]= {\cal J}^A_B l^B \Gamma^A $. Hence, for any 
$l\in L_\Phi$,  
\[
0={\cal J}\cdot l\cdot \Phi= [ {\cal J }\cdot, l\cdot ]\Phi+l\cdot {\cal J }\cdot \Phi=
l\cdot {\cal J }\cdot \Phi\ .
\]
Since two forms with the same annihilator are proportional, ${\cal J }\cdot\Phi= \alpha\Phi$ for some $\alpha$. In a similar way, one sees that elements of $L_\Phi$ raise the eigenvalue of ${\cal J }\cdot$ by $i$, and elements
of $\bar L_\Phi$ lower it by $-i$. Since $U_{-3}$ is obtained by acting on $\Phi$
 with all six creators, it is generated by the conjugate Clifford vacuum $\bar \Phi$; hence  ${\cal J }\cdot\bar \Phi=(\alpha-6i)\bar \Phi $. ${\cal J }\cdot $ is an antisymmetric real matrix (due to $\Gamma^{AB}$) as an operator on differential forms, 
and hence can only have purely imaginary eigenvalues; it follows that $\alpha=3i$, and
then the result we wanted: 
\begin{equation}\label{eq:Jeig}
{\cal J }\cdot U_{\cal J }^k= k\,i \,U_{\cal J }^k .	
\end{equation}

Given two compatible pure spinors $\Phi_\pm$, because of (\ref{eq:commute}) one 
can decompose the bundle of differential forms in subbundles that are simultaneous eigenspaces for the actions of both ${\cal J }_\pm$. 
We can arrange this decomposition for the forms in a Hodge--like diamond \cite{michelsohn,gualtieri-hodge}. This basis is described explicitly using the relation between differential  forms and bispinors in (A.20) of \cite{gmpt3}. One can use that bispinor picture to compute the action of 
$ *\lambda$  on that diamond by using 
\begin{equation}
	* \lambda (C)= i\gamma\, C  
\end{equation}
in which we have identified, for notational simplicity, differential forms and bispinors. 
 We summarize the results for the eigenvalues of all these operators here:
\begin{equation}\label{eq:eigenvalues}
	\begin{picture}(100,100)(180,0)
		\put(-10,60){$({\cal J_+ }\cdot\,,{\cal J}_-\cdot)\ :$}
		\put(0,0){$
\begin{array}{c}\vspace{.1cm}
	(3i,0)\\ \vspace{.1cm}
	(2i,i)  \hspace{1cm} (2i,-i) \\ 
	(i,2i) \hspace{1cm}(i,0) 
	\hspace{1cm} (i,-2i)\\
	(0,3i) \hspace{1.3cm} (0,i)
	\hspace{1cm} (0,-i)
	\hspace{1cm}(0,-3i)\\
	 (-i,2i) \hspace{1cm} (-i,0)
	\hspace{1cm}(-i,-2i)\\
	(-2i,i) \hspace{1cm} (-2i,-i)\\
	 (-3i,0)\\
	  \end{array}$} 
	 	\put(270,60){$*\lambda\ :$}
	 	\put(280,0){$
	\begin{array}{c}\vspace{.1cm}
		i\\ \vspace{.1cm}
		i  \hspace{1cm} -i \\ 
		i \hspace{1cm}-i 
		\hspace{1.2cm} i\\
		i \hspace{1.1cm} -i
		\hspace{1.2cm} i
		\hspace{1cm}-i\\
		 -i \hspace{1.2cm} i
		\hspace{.9cm}-i\\
		i \hspace{.9cm} -i\\
		 -i\\
	  \end{array}\ $ }
		\put(420,-40){.}
			\end{picture}\vspace{2cm}
\end{equation}
One should stress that this is {\it not} the usual Hodge diamond: the forms in each
entry do not have definite $(p,q)$--degree. For example, the entry at the very top is
not 1 but $\Phi_+$. 


\subsection{New supersymmetry equation}\label{sub:new} 

We are now ready to get rid of the $*$ in (\ref{eq:gmpt2i}). We will present the
explicit computation in the type IIB case. The IIA case works essentially in the same way; we 
will  give the result for it at the end of this subsection.
 
First of all, we expand
the RR flux $F=\sum_{p,q}F_{pq}$ in the Hodge basis described in the previous subsection. As shown in 
\cite{gmpt2}, (\ref{eq:gmpt2},\ref{eq:gmpt2i}) imply $F_{30}=F_{03}=0$. 
Then, using (\ref{eq:eigenvalues}), we have
\begin{equation}\label{eq:slF}
	*\lambda(F)= -{\cal J }_- \cdot F+ 2 *\lambda(F_{10}+F_{01}+F_{32}+F_{23})\ .
\end{equation}
One can now use again (\ref{eq:gmpt2},\ref{eq:gmpt2i}) to show that 
\begin{equation}\label{eq:1032}
	\frac i8 e^{3A}*\lambda(F_{10}+F_{01})=dA\wedge \Phi_+\ , \qquad
	\frac i8 e^{3A}*\lambda(F_{32}+F_{23})=-dA\wedge \bar\Phi_+\ . 
\end{equation}
Using (\ref{eq:slF}) and (\ref{eq:1032}), after some manipulations one gets 
\begin{equation}
	\label{eq:FdJB}
	F=-8d_H^{{\cal J }_-}(e^{-3A}{\rm Im}\Phi_+)\ , 
	\end{equation}
where we have defined the operator 
\begin{equation}\label{eq:dJ}
	d^{{\cal J }}_H\equiv[d_H, {\cal J }\cdot]\ .
\end{equation}
This is a generalization of the $d^c=i(\bar\del - \del)$ operator in complex geometry, 
where $\del$ is the Dolbeault operator. We will see this in more detail in section
\ref{sub:the_dd_}; see formula (\ref{eq:dc}). Although we will set $H=0$ in that section, 
there is no problem in generalizing that discussion to $H \neq 0 $, and in defining a twisted
generalized Dolbeault operator so that 
\begin{equation}\label{eq:dcH}
	d_H^{{\cal J }}= i (\bar\del_H^{\cal J }-\del_H^{\cal J })\ .
\end{equation}

The computation in type IIA can be obtained by exchanging $\Phi_+\leftrightarrow \Phi_-$
and by substituting in the above $F_{i,j}\leftrightarrow F_{3-i,j}$. In the notation 
of (\ref{eq:formparity}), we have
\begin{equation}
	\label{eq:FdJ}
	F=-8d_H^{{\cal J }_1}(e^{-3A}{\rm Im}\Phi_2)\ , \qquad d_H^{{\cal J}_1}\equiv[d_H, {\cal J }_1\cdot]\  
\end{equation}
for both theories. One can rewrite it using (\ref{eq:dcH}) so that it looks like (\ref{eq:main}).
It also follows from (\ref{eq:bianchi}) that
\begin{equation}
	\label{eq:deldelbar}
	\delta= -8d_H d_H^{{\cal J }_1}(e^{-3A}{\rm Im}\Phi_2)
\end{equation}

This is the result we were looking for: the Hodge star has been eliminated in favor 
of an action ${\cal J}\cdot$. As we noticed in the previous subsection, this action 
{\it does not} require the metric for its definition.  Both ${\cal J}\cdot$ and (\ref{eq:dJ}) appear naturally in generalized complex geometry, as we will discuss in section \ref{sec:moduli}. In the particular case of O5 solutions, such a phenomenon
was noticed in \cite{gmpt2} (compare eq.~(5.4) there with (\ref{eq:deldelbar}) here). In that case, $d d^{{\cal J}_1}$ reduces to the ordinary
$\del \bar \del$; the equations become essentially the S-dual of the equations 
for ${\cal N}=1$ with NS flux only in type II, that differ from the system found in 
\cite{strominger} only because of the different source equation.

\subsection{The AdS$_4$ case} 
\label{sub:ads}

Before we move on, let me comment on the AdS$_4$ generalization
of (\ref{eq:FdJB}). First of all, the analogue of (\ref{eq:gmpt2},\ref{eq:gmpt2i}) read
\begin{eqnarray}
	\label{eq:gmpt2AdS}
	\begin{array}{cc}\vspace{.3cm}
		d_H \Phi_1=-2 \mu\,e^{-A} {\rm Re} \Phi_2 \  \hspace{1cm}& \left(\mu\equiv\sqrt{-\frac\Lambda 3}\right) \ ,   \\
		d_H(e^A{\rm Im} \Phi_2)=-3 \mu\,{\rm Im} \Phi_1+\frac{e^{4A}}8 *\lambda(F) \ ,\qquad &  ||\Phi_1||=||\Phi_2||=
			\frac{e^{3A-\phi}}{\sqrt{8}}	
	\end{array}
	\end{eqnarray}
where $\Lambda$ is the cosmological constant of AdS$_4$. Notice
that an equation for  $ {\rm Re} \Phi_2$ would also be present, 
but it is now implied by the equation for $\Phi_1$. 

As in the Minkowski case, we would like to reformulate the equation
for ${\rm Im} \Phi_2$ in such a way as to not contain $*$ anymore. 
In this case, it is more convenient to follow a 
procedure slightly different from the one outlined in the 
previous subsection, for reasons that will be clear shortly. 
Looking back 
at (\ref{eq:eigenvalues}), we can see that\footnote{This formula was first pointed out
to me by M.~Gualtieri.}
\begin{equation}
	*\lambda=-e^{\frac\pi 2 ({\cal J}_1 + {\cal J}_2)\cdot}\ ,
\end{equation}
a formula that one could also check using the bispinor formalism explained for example in \cite{gmpt3}, by noticing that ordinary gamma matrices acting from the left anticommute
with $* \lambda$ and those acting from the right commute with it. 
One can now write $* \lambda F= -e^{\frac{\pi}2 {\cal J}_-\cdot}F
+ 2i(F_{10}-F_{01}+F_{32}-F_{23})$, similarly to (\ref{eq:slF}). 
With a few manipulations one gets (for IIB) 
\begin{equation}
	\label{eq:FdJAdS}
	-\frac18 F=d_H^{{\cal J }_-}(e^{-3A}{\rm Im}\Phi_+)+
	3e^{-4A}\mu\,{\rm Re} \Phi_- .
\end{equation}
$d^{{\cal J}_-}_H$ is now defined by (\ref{eq:jdj}) below. As shown
in appendix \ref{sec:review}, in the Minkowski case this definition is equivalent to (\ref{eq:dJ}). (In the previous subsection we chose
to avoid the complicated--looking $e^{\frac{\pi}2 {\cal J}_-\cdot}$.)
In the AdS case, however, since ${\cal J}_-$ is not integrable, 
(\ref{eq:jdj}) and (\ref{eq:dJ}) are not equivalent; the first 
operator squares to zero, the second does not. 




\section{Moduli}\label{sec:moduli} 

In this section we will apply our reformulation of the supersymmetry equation to the 
problem of counting moduli of a supersymmetric solution. We will do so in the
case $H=0$, for simplicity.

We summarize here the system of equations whose moduli problem we want to solve: 
\begin{equation}\label{eq:listH0}
\begin{array}{lc}\vspace{.2cm}
	d \Phi_1=0 \ , \qquad\qquad& d (e^{-A}{\rm Re}\Phi_2)=0 \ ,   \\
	\delta= -8d d^{{\cal J }_1}(e^{-3A}{\rm Im}\Phi_2) \ ,\qquad\qquad&  ||\Phi_1||=||\Phi_2||\ ,
\end{array}
\end{equation}
where again $\Phi_{1,2}$ are compatible pure spinors, reviewed in section \ref{sec:review}. 
$\delta$ will be taken to be fixed for all of this section, but for a quick comment at the
end of section \ref{sub:orientifolds} (see footnote \ref{foot:delta}).
In the compact case, the source $\delta$ will include an orientifold plane; then it
is necessary to add the condition that the pure spinors transform appropriately 
under the orientifold action $\sigma$. For example, for an O5 solution we have 
$\sigma_{O5}(\Phi_+)=\lambda(\bar\Phi_+)$ and $ \sigma_{O5}(\Phi_-)=-\lambda(\Phi_-)$. 
The complete list is found in \cite[Table 4.2]{benmachiche-grimm}. We will ignore these
restrictions for most of this section, and comment on them in section \ref{sub:orientifolds}. 

Notice also that the dilaton is determined via $||\Phi_1||=
\frac{e^{3A-\phi}}{\sqrt{8}}$; hence it is irrelevant for the moduli problem, since it
is determined by the geometrical data and by the warping $A$. 

So we have in (\ref{eq:listH0}) the supersymmetry equations and the relevant Bianchi
identities. This is all we need to satisfy to have a supersymmetric vacuum. 

Let us start by some general remarks about the structure of the system (\ref{eq:listH0}). 
The first thing to notice is that $\Phi_+$ and $\Phi_-$ are coupled through the third
equation, $\delta= -8d d^{{\cal J }_1}(e^{-3A}{\rm Im}\Phi_2)$, through their compatibility condition, (\ref{eq:commute}) (or (\ref{eq:comp})) and through the equal norm condition (last in (\ref{eq:listH0})). 

Let us compare this to the deformation problem for Calabi--Yau manifolds. The forms $J$ and $\Omega$ are still coupled by compatibility and equal norm (which in this case read $J\wedge \Omega=0$ and $i \Omega\bar \Omega=\frac 43 J^3$); but the differential conditions, $dJ=0=d \Omega$, are decoupled. As we know, it turns out that
the algebraic compatibility conditions are not so important. The deformation $\delta \Omega$ contains representations 6, $\bar 3$ and 1 of ${\rm SU}(3)$; the deformation $\delta J$ contains representations 3, 8 and 1. The compatibility conditions relate
the two 3's (which then disappear given that $H^{(1,0)}=0$) and the two 1's. As a result, 
it is possible to express the deformation problem purely in terms of cohomology. 

This example should make us confident that the algebraic conditions are not dangerous, 
but also tells us that the appearance of both pure spinors in the differential equation
$\delta= -8d d^{{\cal J }_1}(e^{-3A}{\rm Im}\Phi_2)$ leads us to uncharted territory. 

\subsection{Real and imaginary part of a pure spinor}\label{sub:real} 

Before we go into that, we will review a few algebraic facts about pure spinors that we
will later need to apply to both $\Phi_1$ and $\Phi_2$. 

The fact that we will mainly need is that a pure spinor $\Phi$ is determined by its real part. If we define $\rho\equiv {\rm Re}\Phi$ and $\hat \rho={\rm Im}\Phi$, from (\ref{eq:Jeig})
we have
\begin{equation}\label{eq:imre}
\hat \rho= -\frac13 {\cal J}\cdot \rho	
\end{equation}
where ${\cal J}$ can also be defined by $\rho$ via \cite{hitchin-gcy}
\begin{equation}\label{eq:JQ}
	{\cal J}_{AB}=  \frac{{\cal Q}_{AB}}{\sqrt{-{\rm Tr}{\cal Q}^2/12}} \ ,  \qquad 
	{\cal Q}_{AB}=(\rho, \Gamma_{AB}\rho)	\ .
\end{equation} 

Conversely, one can ask when a real three--form $\rho$ can be the real
part of a pure spinor, 
\begin{equation}\label{eq:rho}
\rho={\rm Re}\Phi	\ .
\end{equation}
This is a (pointwise) purely algebraic
problem, and it can be shown that it can be solved exactly \cite{hitchin-gcy, hitchin-67} 
when 
${\rm Tr}{\cal Q}^2<0$ (as should be the case for ${\cal J}_{AB}$ to be a generalized complex structure, since it should square to $-1$)\footnote{This was given a physical interpretation in terms
of black hole entropy in \cite{hmt}.}. If that condition is met ($\rho$ is then called {\it stable}), then the imaginary part is determined by (\ref{eq:imre}) and (\ref{eq:JQ}). 
Let us denote ${\rm Im}\Phi$ determined in this way from $\rho= {\rm Re}\Phi$ by $\hat \rho$, 
so that
\begin{equation}
	\Phi= \rho+ i \hat \rho \ .
\end{equation} 

The norm $(\Phi,\bar \Phi)= -2i (\rho,\hat \rho)$ of the pure spinor so determined 
is also proportional to $\sqrt{{\cal Q}^2}$: 
\begin{equation}\label{eq:s}
	s(\rho)\equiv \sqrt{-\frac{{\rm Tr}{\cal Q}^2}{12}}=\frac1{12}\frac{{\cal Q}^{AB}{\cal Q}_{AB}}s=
	\frac1{6s}(\rho, {\cal Q}\cdot \rho)=-\frac12 (\rho, \hat \rho)\ 
\end{equation}
where we have used (\ref{eq:imre}) and (\ref{eq:JQ}) again. For us, the main use of this identity
is that it is now easy to vary the norm of $\Phi$:
\begin{equation}\label{eq:deltaS}
	\delta (\rho, \hat \rho)= 2(\delta \rho, \hat \rho)\ .
\end{equation}
It is also possible to use this to show that \cite{hitchin-gcy} 
\begin{equation}
 	\delta \hat \rho= 	J^{\rm Hit} \delta\rho\ 
\end{equation}
where $J^{\rm Hit}$ is a complex structure associated to $\Phi$ (and hence to $\rho$)
defined by
\begin{equation}
	J^{\rm Hit} \omega = 
	\begin{array}{cc}\vspace{.2cm}
		-i \omega & \qquad {\rm if} \ \omega \in U_{\cal J}^3, \ U_{\cal J}^1\\
		i \omega & \qquad {\rm if} \ \omega \in U_{\cal J}^{-3}, \ U_{\cal J}^{-1}
	\end{array}\ . 
\end{equation}
Using (\ref{eq:Jeig}), one can relate $J^{\rm Hit}$ to the action ${\cal J}\cdot$:
\begin{equation}
	\label{eq:jhj}
	J^{\rm Hit} \omega= -{\cal J}\cdot \omega-2 \frac{{\rm Re}((\omega,\bar \Phi)\Phi)}{(\rho,\hat \rho)}
	= -{\cal J}\cdot \omega
	-2\frac{(\omega,\rho)}{(\rho,\hat\rho)}\rho-2\frac{(\omega,\hat\rho)}{(\rho,\hat \rho)}\hat \rho\ .
\end{equation}

\subsection{Non--integrable moduli}\label{sub:non} 

In this subsection, we will fix $\Phi_1$ and consider deformations of $\Phi_2$; we call these
``non integrable" because $\Phi_2$ (as opposed to $\Phi_1$) is not closed, and hence
the corresponding generalized complex structure ${\cal J}_2$ is not integrable (as we will
review in detail in section \ref{sub:gcmoduli}, for example around (\ref{eq:W5})).

For this particular subset of deformations, 
let us now write the deformation equations in terms of 
$\rho_2=e^{-A}{\rm Re}\Phi_2$, $\hat \rho_2=e^{-A}{\rm Im }\Phi_2$ and their variations:
\begin{equation}\label{eq:nonintmod}
	d (\delta \rho_2) =0 \ , \qquad d d^{{\cal J}_1}\delta(e^{-2A}\hat \rho_2)= 0\ , \qquad
	\delta( e^{2A}(\rho_2,\hat \rho_2))=0 \ , \qquad {\cal J}_1\cdot \delta \rho_2 =0 \ ;
\end{equation}
the first three equations come from  and the last comes from (\ref{eq:comp}). Notice
that the stability condition guaranteeing that $\rho$ can be the real part of a pure
spinor is open; hence (\ref{eq:nonintmod}) is all we need -- $\delta \rho_2$
is an arbitrary form, we need not specify any purity. Let us also recall that $\hat \rho_2$ is 
a function of $\rho_2 $ given by (\ref{eq:imre}) and (\ref{eq:JQ}). 

From the third equation in (\ref{eq:nonintmod}) we get, using (\ref{eq:deltaS}), that 
\begin{equation}
	\delta A = -\frac{(\delta \rho_2, \rho_2)}{(\rho_2,\hat \rho_2)}\ . 
\end{equation}
We can now use (\ref{eq:deltaS}) and (\ref{eq:jhj}) to rewrite the second equation 
as
\begin{equation}\label{eq:ddJmod}
	dd^{{\cal J}_1}e^{-2A}({\cal J}_2\cdot+ 2 F_{\rho_2}) \delta \rho_2=0 \ 
\end{equation}
where we have defined
\begin{equation}
	F_{\rho_2} \omega\equiv \frac{(\omega, \rho_2)}{(\rho_2,\hat \rho_2)}\rho_2  \ ,
\end{equation}
which is very similar to a projector on $\rho_2$: it takes the component over $\hat\rho_2$
and multiplies by $\rho_2$. 

It remains to quotient by symmetries. These are diffeomorphisms and $B$ field gauge
transformations $B\to B+ d \xi$. One can combine such a $\xi$ and a vector $v$ generating
a diffeomorphism in an element $A \in$ \tts, collectively acting as
\begin{equation}
	\omega\to \omega+ d(A\cdot \omega ) 
\end{equation} 
on any closed form $\omega$. 
However, $A$ should be chosen in such a way that
$\Phi_1$ is not affected by them, since this is the assumption in this section. 
Hence one should take $d(A\cdot \Phi_1)=0$. If one does so, one can show that 
$\delta \rho_2 =d(A\cdot \rho_2)$ satisfies the conditions in (\ref{eq:nonintmod}).

Putting all together, we have that the moduli for $\Phi_2$ should be in 
\begin{equation}
	\label{eq:nonintcoh}
	\frac{U^0_{{\cal J}_1}\,\cap \,{\rm Ker}(d)\, \cap {\rm Ker}\Big(dd^{{\cal J}_1}e^{-2A}
	({\cal J}_2\cdot + 2 F_{\rho_2})\Big)}{\{ d(A\cdot \rho_2)\, |\, d(A\cdot \Phi_1)=0\}}\ .
\end{equation}
This result is very similar to the description of moduli for heterotic compactifications in 
\cite{becker-tseng-yau}. This resemblance is looking more evident when one takes the particular case 
of a solution of O5 type; as noted earlier, below (\ref{eq:deldelbar}), this case is
dual to a type II relative of the heterotic equations in \cite{strominger} whose moduli 
are studied in \cite{becker-tseng-yau}.

The next step would be to include deformations of $\Phi_1$. We will not be able to analyze
conclusively the whole system. In section \ref{sub:gcmoduli}, we will review what 
is known about deformations
of the condition $d \Phi_1=0 $, the first equation in (\ref{eq:listH0}). Then, in section 
\ref{sub:are}, we will sketch an approach to the full system. 

\subsection{The $dd^{\cal J}$--lemma}\label{sub:the_dd_} 

The equation $d \Phi_1 =0$, whose moduli we want to review in section \ref{sub:gcmoduli},  implies that  the manifold is
generalized complex. If it also satisfies the so--called $dd^{\cal J}_-$ lemma, its
moduli problem simplifies somehow. In this subsection, we review what this lemma means. 

We should first have a closer look at the operator $d^{\cal J}\equiv [d,{\cal J}]$ we defined 
in the previous section. As we did in section \ref{sub:the_action},
we can now look at what this operator reduces to in the
two canonical special cases, namely for a ${\cal J}$ induced by a complex or by a 
symplectic structure. 

For ${\cal J}={{I \ \ \ 0}\choose {\ 0\ \ -I^t}}$, with $I$ a complex structure, $d^{\cal J}$
reduces to the operator $d^c=i(\bar\del -\del)$, with $\del$ the usual Dolbeault operator. In fact, one can define the generalizations of 
$\del$ and $\bar\del$ for a more general ${\cal J}$ \cite{gualtieri}: for a form $\phi_k \in U^k_{\cal J}$, decompose $d \phi_k$ in a part in $U^{k+1}_{\cal J}$ (and call it 
$\del^{\cal J}(\phi_k)$) and a part in $U^{k-1}_{\cal J}$ (call it $\bar\del^{\cal J}(\phi_k)$). 
It is then still true that 
\begin{equation}\label{eq:dc}
	d^{\cal J}= i(\bar\del^{\cal J}-\del^{\cal J})\ .
\end{equation}

In the symplectic case (${\cal J}={{\ 0\ \ -J^{-1}}\choose{\!\!\!J\ \ \ \ 0}}$, with $J$ the 
symplectic two--form)
the operator $d^{\cal J}$ can also be easily computed. We have $[d,J\wedge]=(dJ)\wedge = 0$; and
$J^{-1}\llcorner$ is nothing but the operator called $\Lambda$ in symplectic
(and, in particular, K\"ahler) geometry. Notice again that $J^{-1}$ is just the 
inverse of $J$; we need no metric to raise the indices. Then $d^{\cal J}$ is equal to $[d,\Lambda]$, which is
traditionally called $\delta$ and studied in symplectic geometry (for example 
\cite{koszul,brylinski,yan,merkulov}; see chapter 5 in \cite{cavalcanti} for a review). 
The idea is to define an alternative Hodge theory based on the antisymmetric form $J$
rather than on the metric. 

Notice that $d^{{\cal J}}$ is a differential, meaning that it squares to zero; and so are
the generalized Dolbeault operator $\del^{{\cal J}}$ and its conjugate. 

Let us now discuss the ``lemma" in the title of this subsection. This is actually a
{\it property} that a generalized complex manifold can or cannot have. 
A generalized complex structure ${\cal J}$ satisfies the generalized $d d^{\cal J}$ lemma 
if any form which is $d$--exact and $d^{\cal J}$--closed is also $dd^{\cal J}$--exact:
\begin{equation}\label{eq:lemma}
	d^{\cal J}(d \alpha)=0 \ \Rightarrow \ d \alpha= d d^{\cal J}\beta \ . 
\end{equation}
From (\ref{eq:jdj}) below, one can see that this also implies that any form which is
$d^{\cal J}$--exact and $d$--closed is $dd^{\cal J}$--exact. 

The original lemma says that the property (\ref{eq:lemma}) is valid on \ka\ manifolds. 
Gualtieri \cite{gualtieri-hodge} showed that it is also valid for each of the two 
generalized complex structures ${\cal J}_\pm$ that define together a generalized 
\ka\ structure. 

The property can, however, be valid without the manifold being \ka\ or generalized \ka. 
For example \cite{deligne-griffiths-morgan-sullivan}, complex manifolds birational to \ka\ manifolds satisfy (\ref{eq:lemma}). 
In the symplectic case, the $dd^{\cal J}$--lemma is equivalent to the so--called
Lefschetz property, as reviewed in \cite{cavalcanti}. 


\subsection{Generalized complex moduli}\label{sub:gcmoduli} 

The deformation problem for the equation 
\begin{equation}
	d \Phi=0
\end{equation}
has been studied from several points of view \cite{hitchin-gcy,gualtieri,li}. In this subsection 
we will review the different approaches. Even if this can only be relevant to the $\Phi_1$ in our
equations (\ref{eq:listH0}), in this subsection we will drop the subscript ${}_1$. 

As a warm--up, let us derive the Kodaira--Spencer equation from a slightly unusual perspective.
Let us start from a complex structure $I$ with holomorphically trivial canonical bundle $K$.
This means that there is global holomorphic section $\Omega$ of $K$, $d \Omega=0 $. To deform 
this, we can use a tensor $\mu$ whose index structure is $\mu^i{}_{\bar j}$. This acts as 
$\mu\cdot=
\mu^i{}_{\bar j}(dz^{\bar j}\wedge)\iota_{\del_i}$, namely
by contracting the upper index and wedging the lower index. This is an infinitesimal 
action: its finite counterpart can be defined by the exponential\footnote{The superscript
${}^0$ will denote from now on undeformed quantities.}
\begin{equation}
\Omega^0 \to \Omega= e^{\mu\cdot} \Omega^0\ .	
\end{equation}
Any decomposable three--form $\Omega$ (namely, a form that can locally be
expressed as wedge of three complex one--forms) defines an almost complex structure
$I$; the integrability of $I$ is translated into the condition that $d \Omega$ be of type  $(3,1)$, or in other words
\begin{equation}\label{eq:W5}
d \Omega= W_5 \wedge \Omega 
\end{equation}
for some one--form $W_5$. (This strange name is a legacy of early studies of 
almost hermitian manifolds \cite{gray-hervella}.) In this situation, 
the canonical bundle is already topologically trivial, because a global non--vanishing section $\Omega$ exists; if $W_5$ is exact, it can be reabsorbed by rescaling and the
canonical bundle has a {\it holomorphic} global section -- therefore it is holomorphically
trivial. In the following, we will mostly take $W_5=0$.
  
Now we can decide whether the complex structure $I$
determined by the deformed $\Omega$ is integrable or not, without looking at the Nijenhuis tensor. By the formula 
$e^{-A} B e^A=[B,A]+\frac12 [[B,A],A]+\ldots$ we get
\begin{equation}\label{eq:KS0}
	\del(\mu\cdot \Omega^0)+ \left([\bar\del,\mu\cdot]+\frac12 [[\del,\mu\cdot],\mu\cdot]\right)\Omega^0= 0\ .
\end{equation}
where $\del$ is the undeformed holomorphic exterior derivative. Notice that the two terms
belong to $U^1_{\cal J}$ and $U^3_{\cal J}$ respectively; hence they have to vanish 
separately. (The first term does not have to vanish if we allow a non--zero $W_5$.)
We can now read this in two ways. One is to note that in the second term, which belongs
to $U^3_{\cal J}$, the operator in parenthesis acting on $\Omega^0$ is algebraic 
(it contains no derivatives) and that it only contains creators. Hence we get
\begin{equation}
	\label{eq:KS}
	[\bar\del,\mu\cdot]+\frac12 [[\del,\mu\cdot],\mu\cdot]=0
\end{equation}
which is one possible form of the Kodaira--Spencer equation. We can, alternatively, 
define $\mu' \equiv \mu\cdot \Omega^0$, after which the first term in (\ref{eq:KS0}) becomes
$\del \mu'=0$ and the second term can be rewritten as
\begin{equation}
	\label{eq:KS1}
	\bar\del \mu' + \frac12 \del (\mu^2)'= 0 \ .
\end{equation}

We can now generalize this. Similarly to (\ref{eq:W5}), the condition that a generalized
complex structure ${\cal J}$ be integrable can be read off the corresponding pure spinor
\cite{gualtieri}: 
\begin{equation}
	\label{eq:intPhi}
	d \Phi= A\cdot \Phi
\end{equation}
for some $A \in $ \tts, acting as $A_B \Gamma^B$ -- a combination of a one--form and
a vector, this time. Just as we did for $W_5$, we are going to assume that $A=0$ in 
what follows. In exactly the same way as (\ref{eq:KS}), then, one obtains a 
generalized Kodaira--Spencer
\begin{equation}\label{eq:genKS}
	[\bar\del_{\cal J},l\cdot]+\frac12 [[\del_{\cal J},l\cdot],l\cdot]=0
\end{equation}
where this time $l\cdot$ is the action of an element of $\Lambda^2 \bar L_\Phi$ -- 
a wedge of two creators. In alternative, one can now define $l'\equiv l\cdot \Phi$ and
write (\ref{eq:genKS}) as
\begin{equation}
	\label{eq:genKS1}
	\bar\del l' + \frac12 \del (l^2)'= 0 \ .
\end{equation}
that generalizes (\ref{eq:KS1}). Notice that one also obtains the analogue
of the first term in (\ref{eq:KS0}),
\begin{equation}\label{eq:extra}
	\del(l')=0\ .
\end{equation}
One can avoid this equation if we allow a non--zero $A$ in (\ref{eq:intPhi}). We will see that we can remain within the more restrictive $d \Phi=0$ and satisfy (\ref{eq:extra}) automatically if the $d d^{\cal J}$ lemma is satisfied.

In the complex structure case, in which $\Phi= \Omega$, one can derive the integrability condition in a more usual
way by  computing the Nijenhuis tensor $N(I)$, and without ever referring to $\Omega$.
$N(I)$ is obtained as the imaginary part of the condition $\bar P [Pv, P w]_{\rm Lie}=0$,
where $P=\frac12(1-i I)$ is the holomorphic projector. This condition means that the Lie
bracket of two $(1,0)$ vectors is still $(1,0)$: the distribution of $(1,0)$ is integrable. The analogous condition for generalized complex structures involves, 
not surprisingly, elements of \tts\ instead of vectors, and the Courant bracket on 
\tts\ instead of the Lie bracket. One can, however, bypass all this and express the 
integrability of a generalized complex structure ${\cal J}$ as (see for example \cite{guttenberg}) 
\begin{equation}
	\label{eq:int}
	[[d,{\cal J}\cdot],{\cal J}\cdot]=-d\ .
\end{equation}
This is not completely surprising, given the definition of the Courant bracket as 
a derived bracket: 
\begin{equation}
	\label{eq:Courant}
	[A,B]_{\rm Courant}\equiv \{\{ d,A \},B\}-\{\{ d,B \},A\} \ .
\end{equation}
Notice also that the trivector part (namely, the part in $\Lambda^3 T^*$) of 
(\ref{eq:int}) is just (referring to the block decomposition 
${\cal J}={{I\ \ \ P}\choose{J\ -I^t}}$ we already used in (\ref{eq:Jaction}))
\begin{equation}
	[[d,P],P]=0
\end{equation}
which means that the upper--right block $P$ of a generalized complex structure ${\cal J}$
is always a {\it Poisson} bivector ($[[d,\cdot],\cdot]$ reduces to 
the Schouten bracket on multivector fields). This fact has indeed been found 
by explicit computation in \cite{lmtz} and later also noticed in \cite{abouzaid-boyarchenko,crainic}. 

One can now deform ${\cal J}={\cal J}^0+ \delta {\cal J}$ to obtain another form of
the Kodaira--Spencer equation:
\begin{equation}\label{eq:KSme}
	[d^{\cal J},\delta {\cal J}\cdot]+\frac12 [d,[\delta {\cal J}\cdot,{\cal J}\cdot]]+\frac12
	[[d,\delta {\cal J}\cdot], \delta {\cal J}\cdot]=0\ . 
\end{equation}
Even if this looks more complicated than (\ref{eq:genKS}) or (\ref{eq:genKS1}), 
it is the form which is most appropriate to understanding the deformation of the
$dd^{\cal J}$--lemma to be discussed in appendix \ref{sec:open}. This is because deforming $d^{\cal J}$ contains $\delta {\cal J}$ rather than the $l$ above.\footnote{The relation between the two is complicated
by the fact that $\delta {\cal J}$ is real, whereas in (\ref{eq:genKS}) we have taken
$l$ in $\Lambda^2 \bar L$, hence complex. Alternatively we could have taken $l$ to be real and hence with a part in $\Lambda^2 \bar L$ and a part in $\Lambda^2 L$, 
at the price of complicating the exponential expansion in (\ref{eq:genKS}); this would
have corresponded to transforming ${\cal J}\to e^l {\cal J} e^{-l}$.}

The original derivation of the generalized Kodaira--Spencer (formula
(5.2) in \cite{gualtieri}) is
neither of the two we saw above. It does not use the presence of a pure spinor $\Phi$
(just like our second approach by deforming (\ref{eq:int})), but it gives  the same
result as (\ref{eq:genKS}), once one translates it in terms of Lie algebroids.
The result in \cite{gualtieri} is expressed in terms of the second cohomology 
$H^2_L$ of a Lie algebroid whose underlying bundle is $L_\Phi$, the annihilator bundle for $\Phi$. In the case in which ${\cal J}$ is holomorphically trivial, which is of interest in this paper, this cohomology is isomorphic to the cohomology of $\bar\del_{\cal J}$
\cite[Section 4.4]{gualtieri},\cite[Prop.~4]{li}.\footnote{This fact generalizes
the Gerstenhaber--Schack result \cite{gerstenhaber-schack,kontsevich} that the second Hochschild cohomology of a Calabi--Yau
is given by $H^0(M,\Lambda^2 T)\oplus H^1(M,T)\oplus H^2(M,{\cal O})$.}
The quadratic term in (\ref{eq:genKS}) also coincides with the Schouten bracket $[l,l]$
in \cite{gualtieri}, just because (\ref{eq:Courant}) can be extended to elements of 
$\bar L_\Phi$ with arbitrary degree, as checked in \cite{li}. 

Let us summarize the generalized Kodaira--Spencer approach to deformations of 
the condition $d \Phi=0$. Infinitesimal deformations sit in the second cohomology
of the Lie algebroid associated with $L$
\begin{equation}\label{eq:HKS0}
	H^2_L \ ,
\end{equation}
or equivalently in 
the second cohomology (starting from $\Phi$) of the generalized $\bar\del_{\cal J}$ operator.

There are two problems with the situation so far.  The first is equation (\ref{eq:extra}), that
up until now we have ignored. For the generalized complex condition alone, 
one can actually avoid (\ref{eq:extra}) by allowing
a non--zero $W_5$ in (\ref{eq:W5}) or more generally $A$ in (\ref{eq:intPhi}). 
For the stronger condition $d \Phi=0$, which is the one that appears in (\ref{eq:listH0}), 
we have to ask whether (\ref{eq:extra}) modifies (\ref{eq:HKS0}). The second problem is that
infinitesimal deformation given by (\ref{eq:HKS0}) could be obstructed.

Both these issues are solved by assuming that the manifold satisfies
the $dd^{\cal J}$ is satisfied (reviewed in \ref{sub:the_dd_}). For the first, one
can note that \cite[Theorem 4.2]{cavalcanti} tells us
that every de Rham cohomology class can be taken to be both $d$-- and $d^{\cal J}$--closed. 
Hence, the cohomology (\ref{eq:HKS0}) parameterizing infinitesimal deformations can be rewritten more concretely as 
\begin{equation}\label{eq:HKS}
	H^*(M,\cc)\cap U^1_{\cal J}\ .
\end{equation}

For the second problem, the presence of possible obstructions, let us try to solve the equation perturbatively as usual, writing $l=l_1+l_2+\ldots$. The equations at first order are solved by an element $l_1$ in 
(\ref{eq:HKS}). At second order we have
\begin{equation}\label{eq:sec}
	\del l_2'=0\ , \qquad \del (l_1^2)'+ \bar\del l_2'=0\ .
\end{equation}
We have that $\del (l_1^2)'$ is $\del$--exact, and -- by passing to the Lie algebroid (as seen
for example in \cite[page 12]{li}) that it is $\bar\del$--closed. This implies that
$\del (l_1^2)'= \bar\del \del(\sigma_1)$ for some $\sigma_1$. Now $l_2'=-\del \sigma_1$ solves (\ref{eq:sec}). 
In a similar way one can proceed by induction. To prove that deformations are unobstructed
would now require proving that the series converges. These details were given in \cite{li} 
for the generalized \ka\ case; as we just sketched, providing an order--by--order
solution for $l$ only requires the $d d^{\cal J}$, and one would expect convergence to work 
again with this weaker assumption. 

Rather than trying to fill in these details, however, we will now review 
the Hitchin functional approach \cite{hitchin-67,hitchin-gcy}, that exhibits the unobstructed moduli space more directly. 

First of all we need to remember that a pure spinor $\Phi$ is determined by its real part $\rho$, as reviewed in section \ref{sub:non}.
If we start from a closed $\Phi^0$, deformations of the real part $\rho=\rho + \delta \rho$ which are small enough will keep $\rho$ stable (since stability is an open condition, or in other words it is defined by an inequality). Then the condition $d \Phi=0$ imposes $d(\delta\rho)=0$. 

One can then determine $\hat\rho={\rm Im}\Phi$ via (\ref{eq:imre}) and (\ref{eq:JQ}). A priori this $\hat \rho$ is not necessarily going to be closed. However, one can vary $\rho \to \rho + d\sigma$ and hope that for some $\sigma$ the resulting $\hat \rho(\rho + d \sigma)$ will be also closed. 

This is made precise by introducing \cite{hitchin-67,hitchin-gcy} the functional 
\begin{equation}
	\label{eq:hf}
	S(\rho)=\int s(\rho)\ ,
\end{equation}
where $s(\rho)=-\frac12 (\rho,\hat \rho)$ as in (\ref{eq:s}). 
Using (\ref{eq:deltaS}), we see that the
 extrema of (\ref{eq:hf}) under variations $\rho\to \rho+d(\sigma) $ are given by $\hat \rho$
such that $\int(d \sigma, \hat \rho)=0$ for any $\sigma$, which implies, integrating by parts, 
that 
\begin{equation}
	d\hat \rho=0 \ .
\end{equation}
One can then use the functional (\ref{eq:hf}), along with the implicit function
theorem, to prove that a generalized complex manifold with holomorphically trivial 
canonical bundle (that is, so that $d \Phi=0$) has an unobstructed moduli 
space.\footnote{\label{weaker} What is
called $d d^{\cal J}$ lemma in \cite{hitchin-gcy} is actually the property that $d J^{\rm Hit} d\tau \Rightarrow d \tau \in U^2_{\cal J}\oplus U^{-2}_{\cal J}$. One can show that this property
follows from the $d d^{\cal J}$ lemma in this paper, (\ref{eq:lemma}).} 
More specifically, its moduli space is an open set in $H^{\rm even}$ or $H^{\rm odd}$,
depending on the parity of $\Phi$. Notice that this agrees with the result
(\ref{eq:HKS}) once one takes into account the trivial rescaling of $\Phi$.

The conclusion of this subsection is that deformations of the condition $d \Phi=0 $ alone are
easy: they are given \cite{hitchin-gcy} by an open set in the cohomology (\ref{eq:HKS}).\footnote{As we remarked
at the end of section \ref{sub:the_dd_}, a class of manifolds obeying the $dd^{\cal J}$
lemma is given by complex manifolds birational to \ka\ ones. The moduli space for these
is indeed unobstructed, because it is just isomorphic to a component of the discriminant
locus of a \ka\ manifold. This conclusion then is in nice agreement with the nonobstructedness theorem in \cite{hitchin-gcy}. These manifolds have been used in \cite{ckt} to construct
supersymmetric string vacua.} In particular, there 
are no obstruction to worry about, and there is a natural flat metric on the moduli space. 


\subsection{Comments about the full system}\label{sub:are} 

Let us now come back to (\ref{eq:listH0}). 

In section \ref{sub:non}, and more specifically in 
(\ref{eq:nonintcoh}), we have given a cohomological
description of the moduli given by the pure spinor $\Phi_2$, for fixed $\Phi_1$. 

Then, in section \ref{sub:gcmoduli}, we have reviewed what is known about deformations of the first equation in (\ref{eq:listH0}) alone, in various degrees of generality -- in particular concluding that we have an unobstructed moduli space of dimension (\ref{eq:HKS}) when the $dd^{{\cal J}_1}$ lemma is satisfied. We should stress that this moduli space will not be unaffected by the other
equations in (\ref{eq:listH0}), in particular by the third (that we also rewrite in (\ref{eq:last}) below). In fact, one can immediately see
that some of the moduli in (\ref{eq:HKS}) will be lifted. We know from (\ref{eq:comp}) that
$\rho$ must be in $U^0_{{\cal J}_1}$; we also know that $d d^{{\cal J}_1}$ preserves 
$U^k_{{\cal J}_1}$. This implies that 
\begin{equation}\label{eq:obstr}
	{{\cal J}_1}\cdot \delta =0\ .
\end{equation}
This equation will in general lift some of the moduli of $\Phi_1$. But it might not be the
only obstruction coming from (\ref{eq:last}), even if we keep $\Phi_2$ constant. Even if
we assume that the $dd^{{\cal J}_1}$ lemma holds, it will only follow that $\delta$ is
$d d^{{\cal J}_1}$ of something, not necessarily of $e^{-2A}\hat \rho_2$ as it has to be.
 
More generally, one would now like to put these results together and study the full system and letting both
$\Phi_{1,2}$ vary. Even though we do not have a definite result about this problem, we 
would like to outline an approach here that might reveal itself useful in the future. 

The idea is the following. One would start from the two decoupled equations in (\ref{eq:listH0}),
$d \Phi_1=0 $ and (in the notation of section \ref{sub:non}) $d \rho_2=0$. The moduli of the 
first are, as we saw, described by (\ref{eq:HKS}); the moduli of the second by 
\begin{equation}\label{eq:infcoh}
	\frac{U^0_{{\cal J}_1}\,\cap \,{\rm Ker}(d)}{\{ d(A\cdot \rho_2)\, |\, d(A\cdot \Phi_1)=0\}}\ .
\end{equation}
which is the same as (\ref{eq:nonintcoh}) but without the $dd^{{\cal J}_1}$ part. This is because we have not considered the third equation in (\ref{eq:listH0}), that we rewrite here
for convenience, again with the redefinitions above (\ref{eq:nonintmod}):
\begin{equation}\label{eq:last}
	-8 dd^{{\cal J}_1}(e^{-2A}\hat \rho_2)=\delta\ .
\end{equation}
This equation is now not necessarily satisfied, after deforming both $\Phi_1$ (and hence ${\cal J}_1$)
 and $\rho_2$. However, the cohomology (\ref{eq:infcoh}) is in general infinite--dimensional
(as also remarked in \cite{becker-tseng-yau}), since we are not quotienting by all exact forms. 
A natural idea is that one can vary $\rho$ within this infinite--dimensional space, to find
a solution to (\ref{eq:last}). 

One possible way one would go about showing this is a variation on the Hitchin functional
argument reviewed at the end of section \ref{sub:gcmoduli}. Let us assume that the
$d d^{{\cal J}_1}$ lemma is valid. The functional to consider would be now\footnote{A modification of the Hitchin functional including flux was also considered in \cite{jeschek-witt-shame}.} 
\begin{equation}
	\label{eq:modhf}
	\int ||\Phi_-||^2 (\rho,\hat \rho)^2 - (\rho,\xi)\ 
\end{equation}
where $\xi$ is a form such that $\delta= -8 d d^{{\cal J}_1} \xi$; we know such a form exists
because of the $d d^{{\cal J}_1}$ lemma. 
Rather than varying within the full (\ref{eq:infcoh}), it makes sense to vary among
exact forms $\rho\to \rho+ d \sigma$, with $d \sigma \in U^0_{{\cal J}_1}$. 
Such a $d \sigma$ is also $d^{{\cal J}_1}$ closed. Because of the $dd^{{\cal J}_1}$ lemma, we can always write such a variation as $dd^{{\cal J}_1}\tau$ for some 
$\tau \in U^0_{{\cal J}_1}$ (again thanks to the fact that $dd^{{\cal J}_1}$ preserves $U^k_{{\cal J}_1}$). A critical point would then occur when 
 \begin{equation}\label{eq:critp}
 	\int (d d^{{\cal J}_1} \tau, e^{-2A}\hat \rho -\xi)= \int (\tau, -\delta+ d d^{{\cal J}_1}(e^{-2A }\hat \rho))= 0 \ ,\qquad \forall \tau\in U^0_{{\cal J}_1}\ 
 \end{equation}
where we have used (\ref{eq:deltaS}) and $e^{-2A}= \frac{(\rho,\hat \rho)}{||\Phi_1||}$. 
(\ref{eq:critp}) means that $d d^{{\cal J}_1}(e^{-2A }\hat \rho)- \delta$ has no component in 
$U^0_{{\cal J}_1}$, which in turn means that (\ref{eq:last}) is satisfied, since $\hat \rho$
is in $U^0_{{\cal J}_1}$.

Now we have shown that (\ref{eq:last}) describes critical points of (\ref{eq:modhf}) in a particular class of variations -- namely, $\delta \rho\in \{ d \sigma \in U^0_{{\cal J}_1}\}$
$=\{dd^{{\cal J}_1}\tau\  {\rm such\ that}\ \tau \in U^0_{{\cal J}_1}\}$. We would now need to show that the 
Hessian around such a critical point has no zero modes besides symmetries (those in the denominator of (\ref{eq:infcoh}). This would mean showing that
\begin{equation}\label{eq:thorn}
	d d^{{\cal J}_1} \,J^{\rm Hit}_2\, d d^{{\cal J}_1} \tau = 0 \  \Rightarrow \ d d^{{\cal J}_1}\tau =
	d (A \cdot \rho_2)\ .
\end{equation}
A similar point arose in \cite{hitchin-gcy}; this is where the $d d^{{\cal J}_1}$ lemma was used
(but see footnote \ref{weaker}). Unfortunately showing (\ref{eq:thorn}) is made more difficult
by the fact that ${\cal J}_2$ is not integrable (since $\Phi_2$ is not closed). 
This implies, in particular, that given a form $\omega$ in $U^k_{{\cal J}_2}$, $d \omega$
has components not only in $U^{k\pm 1}_{{\cal J}_2}$, but also 
in $U^{k\pm 3}_{{\cal J}_2}$. In other words, we do not have a decomposition $d= \del_{{\cal J}_2} + \bar\del_{{\cal J}_2}$. We do have a decomposition $d= \del_{{\cal J}_1} + \bar\del_{{\cal J}_1}$, since ${\cal J}_2$ is integrable, but that does not help towards showing
(\ref{eq:thorn}). 

If this technical point could be shown in some situation, then one could reason paraphrasing
\cite{hitchin-gcy} as follows. We are trying to deform a supersymmetric solution, which 
we have seen now it can be thought of as a critical point of (\ref{eq:modhf}). 
If the Hessian around this critical point did not have any zero modes, then the critical
point would survive small enough deformations. Zero modes are dangerous because, upon a 
deformation no matter how small, they could develop a non--zero first derivative (a ``tadpole'')
that would destroy the critical point. The functional (\ref{eq:modhf}) is bound to have at least the 
zero modes generated by the symmetries in the denominator of (\ref{eq:infcoh}). These particular
zero modes, however, cannot be lifted by small deformations. For this reason, if one showed
(\ref{eq:thorn}), one would conclude that there is actually an unobstructed moduli space. 
This would be valid for both deformations of $\Phi_1$ and of $\Phi_2$, but there is an 
extra subtlety for the former. One might worry that assuming 
the $d d^{{\cal J}_1}$ lemma is valid for the undeformed ${\cal J}_1$ does not mean that
it is still valid once we deform it. Fortunately, at least this point can be shown: the 
$d d^{\cal J}$ lemma is in general an open condition (as we argue in 
appendix \ref{sec:open}) hence it is still valid after small enough deformations.
 
Its dimension would be given by the deformation space for $\Phi_1$, which is given by 
(\ref{eq:HKS}), and by the deformation space for $\rho_2= e^{-2A}{\rm Re} \Phi_2$. This is no
longer given by (\ref{eq:infcoh}) now, because we have used up all exact forms to find a 
critical point for (\ref{eq:thorn}). All in all, remembering also (\ref{eq:obstr}),
one would get 
\begin{equation}\label{eq:maximal}
 	(H^*(M,\rr) \cap U^0_{{\cal J}_1})\oplus \Big\{ \delta {\cal J}_1 \in 
H^*(M,\cc)\cap (U^1_{{\cal J}_1}) \ | \ \delta {\cal J}_1 \cdot \delta =0\ \Big\}\ .	
\end{equation}
This space is unfortunately large -- since we have assumed the $d d^{{\cal J}_1}$
 lemma, its dimension is equal to ${\rm dim} H^*(M,\rr)$ minus deformations obstructed
by (\ref{eq:obstr}). This has at least the merit of showing the reasonable result that
there are fewer moduli than in the case of a Calabi--Yau without flux. 

However, (\ref{eq:maximal}) is somehow the worst case scenario. We have assumed 
the $d d^{{\cal J}_1}$ lemma and that (\ref{eq:thorn}) is true. If these conditions are not satisfied, there can definitely be fewer moduli than (\ref{eq:maximal}) would suggest. In general there will be a superpotential in which 
both the moduli in (\ref{eq:nonintmod}) and the moduli in (\ref{eq:HKS}) appear, and that
can in principle lift both. It would be interesting to realize this explicitly in connection 
to studies of effective theories on manifolds with \stt\ structures \cite{grana-louis-waldram}.

The most popular and easy flux compactification is the one originated in \cite{dasgupta-rajesh-sethi,grana-polchinski} and dual to the F--theory backgrounds in \cite{becker2}. All moduli but the overall volume can be lifted by the flux in this case. 
This hence provides a prominent counterexample to the pessimistic (\ref{eq:maximal}), underlining
that it is by no means a general conclusion. 

On the other hand, the moduli due to $\Phi_2$ alone, given in (\ref{eq:nonintcoh}) will always be there. So this is the minimal
number of moduli one can obtain for supersymmetric Minkowski compactifications. Although I do
not know how to compute that cohomology in general, it might be doable in some example; \cite{becker-tseng-yau} looks encouraging in that respect. 
 
The conclusion of this section is that (up to some assumptions discussed above) the number of moduli is between the dimensions
of (\ref{eq:nonintcoh}) and (\ref{eq:maximal}). Up to now we have not paid any attention, however,
to the effect of sources, to which we now turn. 


\subsection{Orientifolds and branes}\label{sub:orientifolds} 

The discussion so far has not taken into account the effect of having orientifolds. 
In the compact case, if we want to stay within simple supergravity, 
sources of negative tension are necessary, as shown in \cite{haridass-dewit-smit,maldacena-nunez}.
Orientifold planes contribute to the source term in (\ref{eq:listH0}) with the appropriate sign
(in the context of the pure spinor formalism, this was detailed in \cite{gmpt3}), but they
also require that the pure spinors and the fields transform appropriately under the projection
that defines the orientifold. In addition to the orientifold planes, there might or might not
appear sources due to branes. We are now going to comment on how these affect the moduli. 

We will start with orientifolds. The transformation laws for the pure spinors have been listed in  \cite[Table 4.2]{benmachiche-grimm}, as we mentioned. The moduli will be consequently restricted to live in some real subspace of the cohomologies we were proposing in the previous subsections. 
One can actually see that the orientifold transformation law will imply (\ref{eq:obstr}) 
as far as the orientifold source is concerned: 
\begin{equation}\label{eq:obstrO}
	{\cal J}_1 \cdot \delta_{\rm O-planes}= 0 \ .
\end{equation}
If branes are also present in $\delta$, their contribution will also satisfy
${\cal J}_1 \cdot \delta_{\rm D-branes}= 0$; these conditions essentially mean that the O--planes
and the D--branes are generalized complex submanifolds (a concept defined in \cite{gualtieri} and
shown to arise from the brane world--volume action in \cite{koerber,martucci-smyth}). 

Branes will also potentially give rise to new deformations in addition to the geometrical ones we have considered so far. As we remarked at the beginning of this section, $\delta$ has been 
taken so far to be given; deforming the brane sourcese changes $\delta$.\footnote{This does not
mean that any $\delta$ is acceptable; if that were so, (\ref{eq:last}) would have no content, 
and one would end up with infinitely many moduli. Even if one were to accept configurations with ``smeared'' branes, $\delta$ is still constrained by having a Dirac--delta Poincar\'e dual
to the orientifold plane.\label{foot:delta}}
Infinitesimal deformations of the generalized complex submanifold condition have been shown in \cite{koerber-martucci-branes} to be parameterized by a Lie algebroid cohomology
on the submanifold. In general, however, these deformations will mix with the geometrical 
deformations: it is not guaranteed that changing $\delta$ in (\ref{eq:last}) keeps the equation
solvable. We will not consider this problem further. 



\section{Prospects for finding new vacua}\label{sec:method} 

In this section we will consider to what extent the simplification of the equations 
we showed in section \ref{sec:susy}, namely (\ref{eq:FdJ}) or (\ref{eq:deldelbar}), 
might make it easier to find concrete examples. 

Right now there exist several examples of Minkowski flux vacua based on non--Calabi--Yau manifolds, but most of them \cite{dasgupta-rajesh-sethi,kachru-schulz-tripathy-trivedi,becker2-fu-tseng-yau} are 
dual to Calabi--Yau vacua. A few non--dual ones are known \cite{gmpt3}, which have been found
in a (commonly used) approximation in which sources\, (including orientifold planes) are treated as smeared. A class of vacua also exists \cite{ckt} that can be found by using four--dimensional
effective supergravity, but which is beyond the reach of ten--dimensional supergravity. 

From this very short list it is clear that finding new {\it classes} of examples, as opposed
to isolated ones, would be important. The expression (\ref{eq:deldelbar}) should
make it easier to find such examples. To show why, let us go back to the original form 
(\ref{eq:gmpt2i}), which we rewrite as
\begin{equation}\label{eq:dagger}
	8 d^\dagger_{A,H}(e^{-2A}{\rm Re}\Phi_2)= - F\ , 
\end{equation}
where the adjoint $d^\dagger_{A,H}$ is the adjoint to $d_H$ with respect with the ``warped inner product'' $\langle \alpha, \beta \rangle \equiv \int e^{3A} \alpha\wedge * \beta$. (\ref{eq:dagger}) might seem to be tailored for using it with Hodge theory. 
Since we know that $d_H F= \delta$ by Bianchi, we would now apply the Hodge decomposition to 
$\delta$; it would follow that $\delta=d_H( d_{A,H}^\dagger G F)$, where $G$ is the Green operator. 
 Then one would seemingly have solved  (\ref{eq:dagger}) by
\begin{equation}
	\label{eq:dreamon}
	e^{-2A}{\rm Re } \Phi_2 = G \delta  \ .
\end{equation} 
The problem with this is that $G$ depends on the metric, and the metric, in turn, depends on 
${\rm Re}\Phi_2$. Hence (\ref{eq:dreamon}), while true, is not a definition for ${\rm Re}\Phi_2$,
but rather a nonlinear equation that it has to satisfy. The same can be said about (\ref{eq:dagger}): ${\rm Re}\Phi_2$ secretly also appears in the definition of ${}^\dagger$. 

Let us now contrast this with the new version of the same equation, in the form (\ref{eq:deldelbar}). We now have
at our disposal a simpler version of ``Hodge theory'', one that does not lead to the problem
we just described. Namely, after having solved $d_H \Phi_-=0$ by choosing our manifold
to be generalized complex (with holomorphically trivial canonical bundle), we can further restrict ourselves to manifolds on which the $d d^{{\cal J}_1}$ lemma is satisfied. As we
commented in section \ref{sub:the_dd_}, we know plenty of such manifolds.
Now we will have to find an orientifold so that its source $\delta$ is exact, or such that
it becomes exact after adding some branes to it. (This is not an assumption; we know $\delta$ has
to be exact anyway, because $dF = \delta$.) Given that $\delta$ is exact, since it also 
has definite degree
under ${\cal J}_1 \cdot$ (see (\ref{eq:obstr})), it follows that it is $d^{{\cal J}_1}$--exact, and hence $d^{{\cal J}_1}$--closed. By the   $d d^{{\cal J}_1}$ lemma, we get that 
\begin{equation}\label{eq:sol?}
	\delta= d d^{{\cal J}_1} \sigma
\end{equation}
for some $\sigma$. This time this solution is genuinely determined by the mathematics, since the operator
$d^{{\cal J}_1}$, as opposed to $d^\dagger_{A,H} $, does not depend on the metric, but only on 
``half of it", namely on $\Phi_1$. 

This is why (\ref{eq:deldelbar}) represents progress. Unfortunately, this does not mean that
finding solutions to the supersymmetry system has now become trivial. 

One problem is that 
the $\sigma$ we have determined in (\ref{eq:sol?}) via the  $d d^{{\cal J}_1}$ lemma is not necessarily worthy of being called ${\rm Re}\Phi_2$. For a form to be the real part of 
a pure spinor, it has to be stable, as we reviewed 
after (\ref{eq:rho})).  
And even if it is stable, the metric it defines together with $\Phi_1$ might
turn out not to be positive definite. Nothing guarantees, a priori, that $\sigma$ determined
by (\ref{eq:sol?}) will turn out to have both features. 
On the bright side, both these problems (stability and positive definiteness of the metric) are regulated by inequalities. So maybe one can still use the $\sigma$ above, for example to 
perturb a known solution. However, there still would remain to be solved $d(e^{-A}{\rm Re}\Phi_1)=0$.

All these problems might become milder if we renounce full generality and we look at a class
of examples. 
Consider for instance O5 solutions with pure spinors of the form $\Phi_+ =e^{iJ}$,
$\Phi_- = \Omega$. The equations in this case specialize to
\begin{equation}\label{eq:O5}
	d \Omega= 0 \ , \qquad i \del\bar\del(e^{-2A}J)= -\delta\ , \qquad d J^2=0\ , \qquad 
	i\Omega\bar \Omega= \frac 43 e^{2A}J^3\ , \qquad J\wedge \Omega=0 \ .
\end{equation} 
The first condition says that the manifold is complex with holomorphically trivial canonical
bundle, $K=0$. Then, in this case we have $d^{{\cal J}_1}=i (\del - \bar \del)$
with $\del$ the ordinary Dolbeault differential (as we remarked in section \ref{sub:the_dd_}). 
Supposing as above that this complex manifold satisfies the $\del \bar \del$ lemma, 
we have $\delta= i\del \bar \del \sigma$ for some $\sigma$ just like in the general case (\ref{eq:sol?}). Now one possible idea would be to look for a $J_0$ so that $\del\bar\del J_0=0$
and $d J_0^2=0$, and deform these with the $\sigma$ obtained above. The condition $\del \bar\del J_0=0$ is sometimes called SKT (strong \ka\ with torsion); solutions are not difficult to 
find, for example by looking for sigma models with $(2,1)$ supersymmetry. 
\footnote{One case in which one would be able to find solutions is the class of spaces considered in \cite{dasgupta-rajesh-sethi,goldstein-prokushkin,fu-yau,becker2-fu-tseng-yau}, namely $T^2$--fibrations over a K3. 
However, these turn out to be just T--dual to more familiar warped K3$\times T^2$ solutions.}
If one were able to solve $dJ^2 =0$ at the same time, one would end up with the last equation in (\ref{eq:O5}), which now is analogous to the Monge--Amp\`ere equation, but for manifolds
which are only complex and not also \ka. Notice that the ``mirror'' of this, namely the
symplectic case, has been studied in \cite{weinkove,tomassini-weinkove-yau}. In perspective,
I think one should be able to reduce the supersymmetry problem just to such a generalization
of the Monge--Amp\`ere equation. 

\bigskip 

{\bf Acknowledgments.} It is a pleasure to thank G.~Cavalcanti, W.--y.~Chuang, 
M.~Gual\-tieri, M.~Gra\~na, N.~Halmagyi, M.~Headrick, S.~Kachru,
P.~Koerber, Y.~Lin, L.~Martucci, R.~Minasian, M.~Petrini, M.~Schulz, L.--S.~Tseng, D.~Tsimpis, B.~Weinkove, for 
discussions and/or correspondence. This work is supported by the DOE under contract DEAC03-76SF00515 and by the NSF under contract 0244728.

\appendix

\section{Some aspects of generalized complex geometry}\label{sec:review} 

Here we will review some aspects of generalized complex geometry that we need in the
main text, some of them well known, some known but never spelled out in the literature. 

One can view a differential form as a spinor for the Clifford algebra ${\rm Cl}(6,6)$, 
whose generators are the gamma matrices $\Gamma^A=\{\iota_{\del_m}, dx^m\wedge\}$ (see also section \ref{sub:the_action}). We call a differential form {\it pure} if its
annihilator $L_\Phi$ has dimension 6. Two notable examples of pure spinors are 
$e^{iJ}$ ($J$ being a symplectic form) and $\Omega$ (a decomposable complex three--form). 

The internal product of two pure spinors is defined to be
\begin{equation}
 \label{defMukai}
( A , B) \,\mathrm{vol} \equiv \Big(A\wedge 
\lambda(B)\Big)_{\rm top} \ , 
\end{equation}
where  $\lambda$ 
is defined by\footnote{Our conventions are $*_6 e^{a_1\ldots a_k}=
\frac1{(6-k)!}\epsilon_{a_{k+1}\ldots a_6}{}^{a_1\ldots a_k}e^{a_{k+1}\ldots a_6}$ 
and $\gamma=-i\gamma^{456789}$.}
\begin{equation}
    \label{eq:lambda}
\lambda(C_k)=(-)^{[\frac k2]} C_k \ ,    
\end{equation}
$k$ being the degree of the form.
In dimension 6, (\ref{defMukai}) is antisymmetric. It is then convenient to define
the norm of $\Phi$ as  $ (\Phi,  \bar\Phi )=
- i ||\Phi||^2 $. For the purposes of ${\cal N}=1$ supersymmetry, it is also important to 
fix the volume form ${\rm vol}$ in (\ref{defMukai}), since that determines the dilaton 
$\phi$ via the
second equation in (\ref{eq:gmpt2i}). The details of how this is done can be found in \cite{gmpt3}.

One can associate to a pure spinor $\Phi$ a generalized complex structure ${\cal J}$ by demanding that the $i$--eigenspace of ${\cal J}$ be $L_\Phi$. 
Then, given two pure spinors $\Phi_{1,2}$, one calls them {\it compatible} if they 
\begin{equation}
	\label{eq:commute}[{\cal J}_1,{\cal J}_2]=0
\end{equation}
and if the metric $-{\cal I}{\cal J}_1 {\cal J}_2$ on \tts\ is positive definite. One can 
see \cite{gualtieri} 
that a pair of compatible pure spinors determine a positive definite metric (and
a $B$ field) on the manifold. 

One can reformulate the condition (\ref{eq:commute}) directly in terms of $\Phi_{1,2}$ by
asking that 
\begin{equation}\label{eq:comp}
\Phi_2 \in U^0_{{\cal J}_1}	\ ;
\end{equation}
 or in other words, that ${\cal J}_1\cdot \Phi_2=0$ (see discussion around (\ref{eq:Jeig})). 

Let us see why (\ref{eq:comp}) is equivalent to (\ref{eq:commute}). If we have $\Phi_{1,2}$
such that their ${\cal J}_{1,2}$ commute, we can apply the construction in section
\ref{sub:the_action} and get a diamond as in (\ref{eq:eigenvalues}). Then (\ref{eq:comp})
is clear from that equation. 

Conversely, suppose $\Phi_{1,2} $ are pure, and $\Phi_2$ satisfies (\ref{eq:comp}). 
For any $l\in L_{\Phi_2}$ (so that $l\cdot \Phi_2=0$) we have
\begin{equation}
	({\cal J}_1\,l)\cdot \Phi_2= [{\cal J}_1\cdot ,l\cdot] \Phi_2= -l \cdot {\cal J}_1\cdot
	\Phi_2=0\ .
\end{equation}
Hence ${\cal J}_1\, l$ is still in $L_{\Phi_2}$. Hence ${\cal J}_1$ sends $L_{\Phi_2}$
to itself; a similar argument shows that it also sends $\bar L_{\Phi_2}$ to itself. 
We already know that ${\cal J}_1 $  is diagonalizable on \tts\ and now we also
know that it is block--diagonal on $L_{\Phi_2}$ and $\bar L_{\Phi_2}$. An eigenvector
of a block--diagonal matrix can always be decomposed as a sum of two eigenvectors, each
having only components along one of the two blocks only. Hence ${\cal J}_1$ is diagonalizable on $L_{\Phi_2}$ and on $\bar L_{\Phi_2}$: so there exist 
$l_i \in L_{\Phi_2}$, $i=1\ldots 6$, so that ${\cal J}_1 l_i= a_i l_i$ (we also know that
the $a_i$ are either $i$ or $-i$). Together with their conjugates, these form a basis
for \tts. Since the $l_i$ are in $L_{\Phi_2}$, we also have that ${\cal J}_2\, l_i=i l_i$.
Hence the $l_i$ and $\bar l_i$ are a basis of eigenvalues for both ${\cal J}_2$ and
${\cal J}_1$, which means that they commute, (\ref{eq:commute}). This is what we
wanted to show.\footnote{Positivity of the metric still needs to be imposed separately. What we have proven is that compatibility is either (\ref{eq:commute}) and positivity, 
or (\ref{eq:comp}) and positivity.}

(\ref{eq:comp}) can in turn be rephrased as saying that $\Phi_2$ does not have
any component in $U^2_{{\cal J}_1}$ or $U^{-2}_{{\cal J}_1}$, or in other words
\begin{equation}
	( \Phi_2, \Gamma_M \Phi_1)= 	( \Phi_2, \Gamma_M \bar\Phi_1)=0\ .
\end{equation}
This also shows that the condition (\ref{eq:comp}) can also be read $\Phi_1 \in U^0_{{\cal J}_2}$. 

Finally, a  piece of information that we need in the main text
is that from the usual formula $e^{-A} B e^A=[B,A]+\frac12 [[B,A],A]+\ldots$ and (\ref{eq:int}) one obtains 
\begin{equation}\label{eq:jdj}
	e^{-\pi/2 {\cal J}\cdot}d \,e^{\pi/2 {\cal J}\cdot}=-e^{\pi/2 {\cal J}\cdot}d \, e^{-\pi/2 {\cal J}\cdot}=d^{\cal J}\ .
\end{equation}
One can also check this formula by using that, when ${\cal J}$ is integrable, one 
can write $d=\del_{\cal J}+ \bar\del_{\cal J}$  and $d^J=i(\del_{\cal J}-\bar\del_{\cal J})$, and by using (\ref{eq:Jeig}).


\section{The $dd^{\cal J }$ lemma is an open condition}\label{sec:open} 
Suppose that a generalized complex structure ${\cal J}^0$ satisfies the $d d^{{\cal J}^0}$ lemma, 
and consider now a small deformation 
\begin{equation}\label{eq:Jser}
{\cal J}= {\cal J}^0+ \delta_1 {\cal J}+ \delta_2 {\cal J}+\ldots 	\ .
\end{equation}
Suppose we now have a form $\alpha$ which is $d$--exact and $d^{\cal J}$--closed:
\begin{equation}\label{eq:cloex}
\alpha=d \beta \ , \qquad d^{\cal J}\alpha=0	\ .
\end{equation}
What we want to prove is that there exists $\gamma$ such that
\begin{equation}\label{eq:ddg}
	\alpha=d \beta= d d^{{\cal J}} \gamma \ .
\end{equation}
Since we have expanded ${\cal J}$ in a power series (\ref{eq:Jser}), we will also expand
$\alpha$ in a series, $\alpha= \alpha_0 + \alpha_1 + \alpha_2 +\ldots$ (and similarly
for $\beta= \beta_0 + \beta_1 + \beta_2 +\ldots$) so that it solves
(\ref{eq:cloex}). 

At zero--th order, we know
that $d^{{\cal J}^0}\alpha_0=d^{{\cal J}^0}d \beta_0= 0$ implies that $d \beta_0= 
dd^{{\cal J}^0} \gamma_0$ for some $\gamma_0$. At first order, then, (\ref{eq:cloex}) implies
\begin{equation}
	\label{eq:cloex1}
	d^{{\cal J}^0}d \beta_1 + [d,\delta_1 {\cal J}]dd^{{\cal J}^0}\gamma_0\ .
\end{equation}
We want to see if using the 
$d d^{{\cal J}^0}$ lemma we can find a solution for (\ref{eq:ddg}), that at first order
reads
\begin{equation}\label{eq:ddg1}
	\alpha_1 = d (d^{{\cal J}^0} \gamma_1 +[d,\delta_1 {\cal J}\cdot] \gamma_0) \ .
\end{equation}
Now, (\ref{eq:KSme}) at first order implies, upon taking its anticommutator with $d$, 
\begin{equation}
	\label{eq:dKSme1}
	\{d^{{\cal J}^0},[d,\delta_1 {\cal J}\cdot]\}=0\ .
\end{equation}
This allows us, in the second term of (\ref{eq:cloex1}), to pull $dd^{{\cal J}_0}$  in front: 
\begin{equation}
	d^{{\cal J}^0}d(\beta_1 -[d,\delta_1 {\cal J}\cdot] \gamma_0)=0\ .
\end{equation}
Using again the $d d^{{\cal J}^0}$ lemma, it follows 
that $d(\beta_1 -[d,\delta_1 {\cal J}\cdot] \gamma_0)= d d^{{\cal J}^0}\sigma_1$
for some $\sigma_1$. But now this means that we can take the solution to (\ref{eq:ddg1}) to be  $\gamma_1 = \sigma_1 $. 

We could now show how assuming a solution at $k$--th order implies one at $(k+1)$--th, but to avoid an orgy of sums and abstract expressions we will actually show that a solution exists at second order, since the latter contains all the essential elements of the former.\footnote{Another thing we will not show here is that the series actually converges.} 
At second order, (\ref{eq:cloex}) now implies
\begin{equation}
	\label{eq:cloex2}
	d^{{\cal J}^0} \alpha_2 + [d,\delta_1 {\cal J}\cdot ] \alpha_1 +[d,\delta_2 {\cal J}\cdot ] \alpha_0 =0 \
\end{equation}
and we want to solve (\ref{eq:ddg}) expanded at second order:
\begin{equation}
	\label{eq:ddg2}
	\alpha_2 = d( d^{{\cal J}^0} \gamma_2 + [d, \delta_1 {\cal J}\cdot] \gamma_1+[d, \delta_2 {\cal J}\cdot] \gamma_0)\ .
\end{equation}
Since we have solved (\ref{eq:ddg1}), we can now rewrite (\ref{eq:cloex2}) as
\begin{equation}
	d^{{\cal J}^0}\alpha_2 + [d,\delta_1 {\cal J}\cdot]d
	(d^{{\cal J}^0}\gamma_1 + [d,\delta_1 {\cal J}\cdot]\gamma_0)+
	[d,\delta_2 {\cal J}\cdot]d d^{{\cal J}^0}\gamma_0=0\ ;
\end{equation}
using now (\ref{eq:dKSme1}) again, we can rewrite this 
as
\begin{equation}\label{eq:big2}
	d^{{\cal J}^0}d( \beta_2 -[d,\delta_1 {\cal J}\cdot]\gamma_1 -[d,\delta_2 {\cal J}\cdot]\gamma_0)-\Big([d,\delta_1 {\cal J}\cdot][d,\delta_1 {\cal J}\cdot]+\{[d,\delta_2 {\cal J}\cdot],d^{{\cal J}^0}\}\Big)d \gamma_0 =0\ .
\end{equation}
The term in brackets acting on $d \gamma_0$ is nothing else than the anticommutator of (\ref{eq:KSme}) with $d$ expanded at second order; hence it vanishes. From the $d d^{{\cal J}^0}$
lemma, (\ref{eq:big2}) now implies that $d(\beta_2 -[d,\delta_1 {\cal J}\cdot]\gamma_1 -[d,\delta_2 {\cal J}\cdot]\gamma_0)= d d^{{\cal J}^0}\sigma_2$. Looking at (\ref{eq:ddg2}), we
see that we can now solve it by simply taking $\gamma_2 = \sigma_2$.


\providecommand{\href}[2]{#2}

\end{document}